\documentclass[letterpaper,twocolumn,10pt]{article}
\usepackage{usenix}

\usepackage{tikz}
\usepackage{amsmath}
\usepackage{booktabs}
\usepackage{multirow}
\usepackage{hyperref}
\usepackage{subcaption}
\usepackage{dblfloatfix}
\usepackage[available,functional,reproduced]{usenixbadges}

\usepackage[]{graphicx,calc}
\newlength\myheight
\newlength\mydepth
\settototalheight\myheight{Xygp}
\newcommand*\inlinegraphics[1]{%
  \settototalheight\myheight{Xygp}%
  \settodepth\mydepth{Xygp}%
  \raisebox{-2.8\mydepth}{\includegraphics[height=1.8\myheight]{#1}}%
}
\usepackage{titling}
\usepackage{xurl}
\usepackage{paralist}
\usepackage{listings}
\microtypecontext{spacing=nonfrench}
\usepackage[ruled, lined, commentsnumbered, longend,vlined]{algorithm2e}
\usepackage{algpseudocode} 
\usepackage{tabularx}
\usepackage{colortbl}
\usepackage{breakurl}
\usepackage{tikz,eso-pic}

\newcommand{\ProjectName}[1]{{\small\textsc{UGCG-Guard}}} 
\newcommand{\PromptName}[1]{{\small\textsc{UGCG-CoT}}} 

\begin{document}

\date{}

\title{\Large \bf  Moderating Illicit Online Image Promotion for Unsafe User Generated Content Games Using Large Vision-Language Models}


\author{
{\rm 
Keyan Guo$^\ast$, Ayush Utkarsh$^\ast$, Wenbo Ding$^\ast$, Isabelle Ondracek$^\ast$},\\
{\rm 
Ziming Zhao$^\diamondsuit$, Guo Freeman$^\dag$, Nishant Vishwamitra$^\ddag$, Hongxin Hu$^\ast$}\\
$^\ast$University at Buffalo, $^\diamondsuit$Northeastern University\\ $^\dag$Clemson University, $^\ddag$The University of Texas at San Antonio, 
}

\maketitle
\begin{abstract}

Online user generated content games (UGCGs) are increasingly popular among children and adolescents for social interaction and more creative online entertainment. However, they pose a heightened risk of exposure to explicit content, raising growing concerns for the online safety of children and adolescents. Despite these concerns, few studies have addressed the issue of illicit image-based promotions of unsafe UGCGs on social media, which can inadvertently attract young users. This challenge arises from the difficulty of obtaining comprehensive training data for UGCG images and the unique nature of these images, which differ from traditional unsafe content. In this work, we take the first step towards studying the threat of illicit promotions of unsafe UGCGs. We collect a real-world dataset comprising 2,924 images that display diverse sexually explicit and violent content used to promote UGCGs by their game creators. Our in-depth studies reveal a new understanding of this problem and the urgent need for automatically flagging illicit UGCG promotions. We additionally create a cutting-edge system, \ProjectName{}, designed to aid social media platforms in effectively identifying images used for illicit UGCG promotions. This system leverages recently introduced large vision-language models~(VLMs) and employs a novel conditional prompting strategy for zero-shot domain adaptation, along with chain-of-thought (CoT) reasoning for contextual identification. \ProjectName{} achieves outstanding results, with an accuracy rate of  94\% in detecting these images used for the illicit promotion of such games in real-world scenarios.

\end{abstract}
\vspace{0.3cm}
\noindent \textbf{Disclaimer}. This manuscript contains discussions and visual representations of sexually explicit and violent content. Reader discretion is strongly advised.

\section{Introduction}

In recent years, online user generated content (UGC) has steadily shifted into the limelight, captivating a widespread audience. 
The sphere of gaming, in particular, has experienced a transformative impact~\cite{genielabs2023ugc}.
Gaming platforms, such as Roblox, have established revenue-sharing models with these UGC creators~\cite{kou2023harmful}. 
This collaborative approach has attracted a multitude of UGC creators to build their UGC games~(UGCGs) and, as a result, has attracted a large number of users, especially children and adolescents.
Data from December 2022 illustrates that 60\% of its user base is under 16 years old, with a substantial 45\% comprising children who are under 13 years old~\cite{statista2022roblox}. 
To attract users, the creators advertise their UGCGs leveraging online social media platforms~\cite{howtoprompt, marketgame, freeprompt,freeadvertise,gameadvertising}, such as X~\cite{X} (formerly Twitter), Reddit~\cite{reddit}, and Discord~\cite{discord}. In particular, X, as a platform that brings together a large number of UGC creators and gamers, is often utilized as the first choice~\cite{howtoprompt}.
However, the surge in user participation has also attracted individuals with malicious intentions, who have proliferated various harmful games with unsafe content, especially sexually explicit imagery and violence~\cite{bbc, dailymail2018outrage, kou2023harmful, MCDONALD2017770, vanacker2012ethical, wolfendale2007my}. 
These games present an unprecedented safety issue to underage users who, often, are ill-prepared to confront or manage such exposures~\cite{bbc, hackernews2019post}. The exposure to explicit content and interactions violates not only ethical norms but also poses significant challenges to their psychological, emotional, and social development~\cite{medium2023darkside, kou2023harmful,vanacker2012ethical}. 

\begin{figure}[t]
    \centering
    \includegraphics[width=\columnwidth]{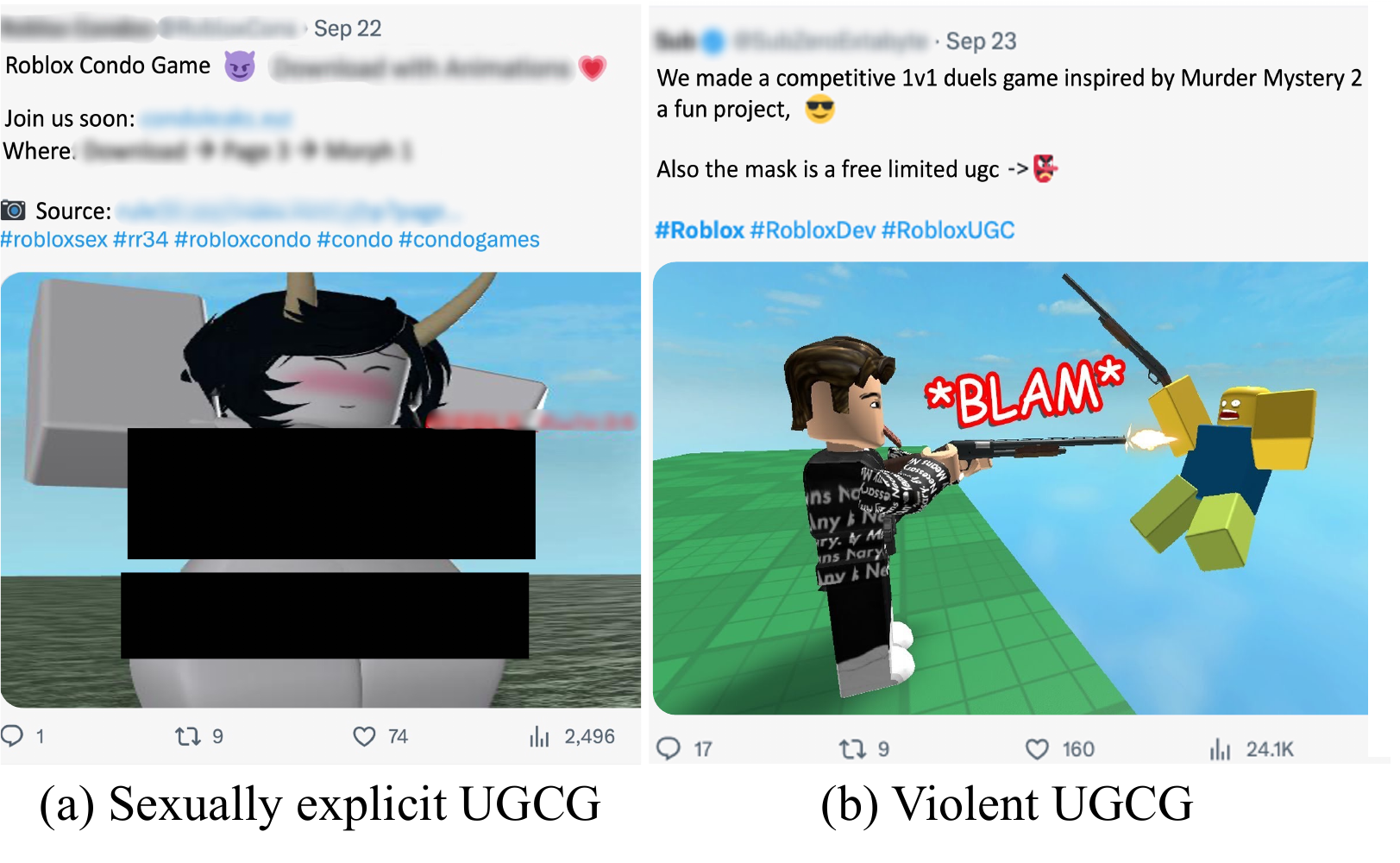}
    \caption{Illicit promotions of unsafe UGCGs on X.}
    \label{fig:sample}
\end{figure}

While moderation during UGCG play is a subject of discussion~\cite{roblox2023safety, robloxsafety2023, blizzard2023conduct,swtor2023conduct},  alarmingly little effort has been made in moderating the \emph{image-based illicit promotion of such UGCGs by malicious creators on social media platforms}, who are resorting to platforms like X to promote their games. 
%
As depicted in Figure~\ref{fig:sample}, the creators share promotional unsafe images of UGCGs to draw a large number of young players to their harmful designs.


Currently, various existing tools, such as Google Cloud Vision API~\cite{google2023vision}, Clarifai~\cite{clarifai2023}, and Amazon Rekognition~\cite{amazon2023rekognition}, utilize artificial intelligence and machine learning~(AI/ML) models for moderating harmful content~\cite{lai2022human}. 
However, there is a concern regarding the effectiveness of these tools in preventing the illicit promotion of unsafe UGCG images. While AI/ML-based systems, such as these detectors, have demonstrated considerable efficacy in identifying traditional unsafe images (\emph{i.e.}, real-world sexually explicit and violent images)~\cite{google2023vision,microsoft2023azure,amazon2023rekognition,clarifai2023,yahoo2016nsfw}, these systems exhibit diminished efficiency when tasked with detecting unsafe images that are used for illicit online promotions of UGCGs. There exist two key problems in flagging such images.
%
\emph{First}, a paramount problem stems from the requirement of extensive training data intrinsic to traditional machine learning models. These models are adept at identifying and classifying conventional unsafe content, such as sexually explicit and violence, through bolstering by large, annotated datasets. However, the acquisition of such large-scale data becomes a formidable task in the context of UGCGs, due to the ambiguous (\emph{i.e.} undefined) nature of content within these virtual worlds, characterized by an eclectic mix of artificially rendered avatars and abstract geometrical representations. 
For example, in Figure~\ref{fig:sample} (a), the avatar is a mix of a female-like character with animal-like horns. 
%
%
\emph{Second}, unlike traditional unsafe images, the UGCG images exhibit a substantial shift in the input domain. 
Traditional AI/ML-based systems are adept at detecting explicit content featuring real human forms. However, UGCGs introduce a complex landscape where there is a transition from real to artificial. The rendered avatars or personas in UGCGs embody a diverse array of forms and contexts, making their classification a complex endeavor. 
While such images are challenging for AI/ML, humans can easily perceive these images due to their contextual knowledge.

In this work, we take the first step towards studying the critical problem of image-based online illicit promotions of UGCGs. 
{We first compile a real-world dataset of images collected from X. This dataset comprises a wide range of illicit online promotional images associated with UGCGs, shared by actual game creators. We collect these images based on keywords derived from a textual analysis of self-reported experiences shared by parents and children on Common Sense Media~\cite{CommonSenseMedia2023} about their experiences with UGCGs.}
{We then conduct a study to explore the characteristics of these image-based illicit online promotions of UGCGs, discovering that the majority of these promotional images are screenshots taken from UGCGs.}
%
%
{We further measure the performance of existing unsafe image detection systems against these illicit promotional images of UGCGs. 
The low success rate indicates that existing systems are severely limited in addressing the challenges presented by these images.}
Our findings underscore the urgent necessity for enhanced detection mechanisms for flagging image-based online illicit promotions of
UGCGs on social media platforms.

Based on our findings, we design \ProjectName{}~\footnote{Our code and datasets are available at \url{https://github.com/UBSec/UGCG-Guard}}, a novel system for flagging images used for the illicit promotion of unsafe UGCGs. 
\ProjectName{} leverages recently introduced advancements of large vision-language models (VLMs) to detect these images, based on a novel conditional prompting strategy designed for zero-shot domain adaptation, which ensures that the model is attuned to the distinct and nuanced characteristics inherent in UGCG's image content, facilitating the flagging of these images without the need of a large dataset, and a chain-of-thought  (CoT) reasoning mechanism~\cite{wei2023chainofthought} for contextual identification of the activities of the personas in these images, enabling \ProjectName{} to discern and respond to the intricate patterns that define illicit promotional images of unsafe UGCGs. Our system achieves a state-of-the-art average accuracy of 94\% in flagging such content.

The key contributions of this paper are as follows:
\begin{itemize}
    \item \textbf{New dataset.}
    We compile a novel, comprehensive dataset consisting of 2,924 images used for unsafe UGCG promotions by the actual game creators on the social media platform X . These images were systematically gathered over the period from the beginning of 2020 to the end of 2022, serving as a valuable resource for analyzing the visual promotional strategies for UGCGs, and enhancing our understanding of how unsafe UGCGs are promoted on such platforms.
    Our dataset will be publicly available for verified researchers to facilitate future research in this area.
    
    \item \textbf{New understanding of unsafe UGCGs and their illicit promotions.}
    We conduct a study to understand the challenges presented by the illicit online image promotion of unsafe UGCGs. 
    We find such promotions utilized inappropriate images, often screenshots taken from UGCGs, for their promotional purposes.
    Our measurement analysis further reveals the severe limitation of existing unsafe image detection systems in identifying unsafe UGCG images, underscoring the urgent need for new moderation approaches to counteract illicit UGCG image promotions.\looseness=-1
    
    \item \textbf{New framework for the moderation of image-based illicit online promotion for unsafe UGCGs.}
    We introduce \ProjectName{}, a state-of-the-art framework to flag image-based illicit promotions of unsafe UGCGs. Rooted in a novel conditional prompting strategy and CoT reasoning approach, \ProjectName{} effectively leverages large VLMs to achieve zero-shot adaption and contextual detection. 
    \ProjectName{} can efficiently distinguish content indicative of image-based illicit promotions of UGCGs, even without prior explicit training on similar content categories.
    
    \item \textbf{Extensive evaluation of \ProjectName{}.}
    Our system's evaluation demonstrates its state-of-the-art average accuracy of 94\%, surpassing existing baseline detectors of unsafe images by 23.7\% to 77.7\% in flagging image-based illicit UGCG promotions. 
    Our experiment also shows that our prompting strategy is significantly efficient, outperforming generalized prompting with an improvement of 64.9\%. 
    In real-world scenarios, our framework successfully identifies and flags image-based illicit promotions of UGCGs, achieving an impressive average F1 score of 0.91.

\end{itemize}

\section{Background and Related Work}
 
\subsection{Online UGCGs and Unsafe Content}
\label{sec:backgroud}
The advent of the internet and social media spurred an interactive cultural evolution characterized by UGC posts online, including images, videos, text, and audio. UGC has significantly influenced gaming, promoting creativity, prolonging game lifespans, and fortifying communities~\cite{genielabs2023ugc}. 
UGCGs like ``Minecraft'' and ``Roblox'' epitomize this shift, transforming players into creators and expanding gameplay possibilities exponentially.
Now, such UGCGs are wildly presented in today's game platforms for social interaction and gameplay.
They allow players to create their own UGC and share it with others~\cite{IGN2023, socialgame2023, genielabs2023ugc}.

However, as the volume of UGCGs proliferates online, there is a corresponding surge in malicious content and activities within these games. 
Online communities in UGCGs are thriving spaces where participants interact through their avatars. While avatars can be vehicles for positive engagement, there is a dark side to this liberty. Some participants exploit this platform to introduce and engage in malicious activities, turning otherwise positive virtual interactions into avenues for undesirable behaviors such as fighting, having sex, and killing others.~\cite{wolfendale2007my, vanacker2012ethical, MCDONALD2017770}.
In addition, a recent study by Kou et al.~\cite{kou2023harmful} revealed another safety concern, the UGC creators misuse the creative latitude offered by platforms like Roblox to introduce harmful designs. 
Examples include games promoting Nazi roleplay and embedding gambling-like mechanisms within these user generated environments. 
The risks are manifold, from the direct introduction of inappropriate content to the more subtle integration of problematic incentive structures within the UGCGs, each posing significant ethical and safety concerns.


\subsection{Content Moderation}
\label{sec:related_work}

Content moderation has been extensively studied as an effective modern mechanism to address issues of toxicity and harassment in online spaces. At a high level, content moderation can be broadly defined as \textit{``the governance mechanisms that structure participation in a community to facilitate cooperation and prevent abuse''} \cite{grimmelmann2015virtues}, and can often be characterized as a series of trade-offs between actions, styles, philosophies, and values based on the context and facilitators of moderation \cite{jiang2022trade}. Current moderation methods for UGCGs include filtering inappropriate or harmful words, filtering personal information, providing a reporting feature, and posting a public Code of Conduct the users of the games in question should obey~\cite{roblox2023safety, blizzard2023conduct,swtor2023conduct}. 
Such content moderation strategies have been widely used in various online contexts but have also demonstrated several limitations. 
For example, TikTok users, who engage with a video-centric social media platform, have devised a strategy to circumvent the platform's algorithmic content moderation by deliberately misspelling prohibited words (e.g., using ``seggs'' instead of ``sex'')\cite{doi:10.1177/20563051231194586}. Cho et al. expand on this topic, discussing a wider array of text-based content moderation evasion tactics in their study\cite{cho-kim-2021-google}.
%
These moderation strategies, however, are primarily text-based.
Although there are effective methods for moderating textual content, there are no comprehensive solutions for moderating illicit image sharing for images used in UGCGs. 

{There exist AI/ML techniques aimed at identifying unsafe visual content, 
including inappropriate images and videos.
Platzer et al. developed a machine learning approach utilizing Support Vector Machines (SVM) for recognizing pornographic images by analyzing distinct image features~\cite{10.1145/2598918.2598920}.
In another study, Yuan et al. explored the real-world issue of illicit online promotions involving pornographic images~\cite{yuan2019stealthy}. 
They introduced a novel deep learning approach that prioritizes regions in an image where sexual content is minimally obscured, aiming to overcome the challenges posed by adversarial sexually explicit images in the real world.} 
{Concerning video content, Tahir et al. highlighted the importance of detecting inappropriate videos within child-centric content-sharing platforms, such as YouTube Kids~\cite{Tahir2019BringingTK}. They collected a dataset of such videos and developed a multimodal model for detection purposes. 
Additionally, another study~\cite{Papadamou2019DisturbedYF} provided a thorough analysis of inappropriate or disturbing videos aimed at toddlers, assembling a large-scale, labeled dataset for this purpose. The researchers trained a deep learning classifier, which yielded a promising detection rate.}
{However, these techniques often rely on evaluating identified features, such as the extent of skin exposure in images and videos, or assessing the ``humanness'' of the subjects depicted~\cite{Hu2007RecognitionOP}.} 
Such criteria can be limiting and may not effectively address the diverse and complex nature of visual content in UGCGs.
A significant gap persists in effectively moderating image content within UGCGs, as the existing AI/ML techniques are not tailored to discern the nuances of unsafe images specific to this domain~\cite{8094233}. Our study underscores this gap and emphasizes the imperative need for specialized moderation methods adept at identifying and mitigating the sharing of unsafe images in UGCGs, ensuring a safer and more inclusive gaming environment.

\subsection{Large Vision-Language Models and Chain-of-thought Reasoning}
\label{sec:LVLMs}

The landscape of multimodal learning, especially in the domain of vision-language multimodal learning, has experienced significant advancements. Pioneering the contemporary discourse, models like CLIP~\cite{radford2021learning} and BLIP~\cite{li2022blip} have marked noteworthy milestones. 
CLIP, for instance, has revolutionized the field by learning visual concepts from natural language descriptions, forging a symbiotic relationship between vision and language components to enhance performance across a multitude of tasks. 
BLIP further accentuates this integration, exemplifying robust performance and versatility in applications ranging from zero-shot to few-shot learning scenarios.
As the discourse evolves, the emergence of large vision-language models heralds a new epoch in multimodal learning. These models, characterized by their incorporation of large language models (LLMs) to navigate vision-language tasks, are driving unprecedented progress.
LLaVA~\cite{liu2023visual}, another significant development, underscores the synergy between visual and linguistic elements, optimizing performance in complex, dynamic environments. 
InstructBLIP~\cite{dai2023instructblip}, building upon the foundational principles of BLIP, integrates instructional learning paradigms to enhance model interpretability and task-specific adaptability. 
The GPT-4Vision~\cite{openai2023gptv} stands as a testament to the ongoing evolution, amalgamating extensive language modeling capacities with intricate visual comprehension, paving the way for a future where the confluence of vision and language is not just integrated but inherently synergistic.
These cutting-edge large VLMs are characterized by their interpretability, adaptability, and enhanced contextual capability, offering many opportunities to tackle intricate and emerging challenges within the vision-language domain.  
In our study, we exploit the capabilities of large VLMs to address a specific issue: the illicit online promotion of unsafe images in UGCGs. These advanced models provide the necessary tools to identify and analyze subtle and complex patterns of unsafe content dissemination effectively.



Chain-of-thought (CoT) reasoning refers to the process where AI models, particularly LLMs, follow a sequence of logical steps to arrive at a conclusion or answer~\cite{wei2023chainofthought}. It involves connecting various information and ideas coherently and logically, akin to a human's natural thought process. This methodology enhances the model's ability to handle complex queries and problems, offering more contextually relevant and nuanced responses.
The large VLMs with such enhanced reasoning capabilities herald a new phase in AI decision-making~\cite{zhang2023multimodal, ge2023chain, vishwamitra2024moderating}. 
Although CoT has been instrumental in elucidating AI decisions~\cite{Huang_2023, zhang2023multimodal}, its application and potential challenges in the specific realm of unsafe content moderation remain to be thoroughly explored and understood.
\section{Threat Model}
%
{In our work, we examine the threat posed by adversaries who exploit social media platforms to promote unsafe UGCGs with inappropriate images, typically screenshots taken from UGCGs. We consider both adversaries and victims as users of these platforms and do not consider promotions outside the social media platforms.}
We do not attribute advanced capabilities to these adversaries nor assume the application of intricate adversarial tactics to evade content moderation. 
Our focus is on the straightforward yet effective strategies these actors employ to pervade digital platforms, exposing vulnerable online users to unsafe UGCGs.
%

In addition, we do not consider in-game content moderation of harmful content in our paper.
Our system specifically targets 
identifying the illicit promotional images for unsafe UGCGs
on social media platforms following a ``soft moderation'' strategy~\cite{paudel2022lambretta}, \emph{i.e.}, providing warnings to users about the potential harmfulness of the {images}, since some platforms may not consider such {images} illegal. 
X and Reddit allow users to flag {inappropriate images} as sensitive or ``NSFW'', which can prevent such content from appearing automatically in timelines, offering a form of user controlled moderation. Platforms such as X, Reddit, and Discord maintain their community guidelines and systems for reporting and eliminating inappropriate content. However, the enforcement of these guidelines can greatly differ across various community spaces, largely relying on the vigilance of community moderators or the effectiveness of automated detection tools.
Despite these mechanisms, inconsistencies in enforcement create loopholes that enable the continued illicit image-based promotion of unsafe UGCGs, highlighting a significant safety challenge.
The persistent threat of illicit UGCG promotions underscores a universal challenge: no social media platform should permit content that risks the safety of its users, particularly minors. Therefore, enhancing content moderation frameworks to address these gaps is crucial in our model, ensuring the digital environment remains safe and conducive to positive interactions.\looseness=-1
%

Furthermore, we focus on illicit promotions of unsafe UGCGs based on images and do not consider the textual data in such posts, since images are the major components that drive promotions due to their visual appeal. 
We also posit that image-based promotions, by virtue of their visual appeal, pose a significant risk, particularly to children and adolescents.


\section{Motivation and Observation}

In this section, we present studies on understanding the nature of unsafe UGCGs in Roblox and their illicit promotions particularly among children, and the challenges in their detection.
%
%
Our study begins with the analysis of real-world, self-reported stories related to unsafe UGCGs. Utilizing the identified keywords from our analysis, we then collect potential UGCG promotional images from social media platforms and engage human annotators to systematically annotate the images. 
{We analyze the challenges of illicit online image promotions for unsafe UGCGs and}
assess the necessity for moderation techniques to flag illicit image promotions associated with unsafe UGCGs. This involves a comprehensive evaluation of the current tools designated for the detection of unsafe images online.\looseness=-1



\subsection{{Data Collection and Annotation}}
\label{sec:data2}


\noindent {\textbf{Hashtag Identification.}
To find effective hashtags for collecting valid UGCG images, we first gathered online discussions with self-reported stories related to unsafe UGCGs from a public Internet resource, Common Sense Media~\cite{CommonSenseMedia2023}.
This platform was chosen for two main reasons: it offers a comprehensive repository of real-world narratives related to UGCG, and it features insights specifically from children and parents, providing valuable perspectives on the effects of unsafe UGCG on younger groups.
By October 2, 2023, we had compiled all Roblox-related discussions from this platform using an automated crawling technique, resulting in 7,081 stories collected. These stories span from July 9, 2009, to October 1, 2023.
Then, a cleaning process ensued, during which we filtered out duplicated stories, those not written in English, and stories comprising fewer than five words to ensure the quality and relevance of the data.
We employed KeyBERT~\cite{keybert} to automatically extract keywords from the collected stories.
Then, we collected illicit promotional images of UGCGs by identifying hashtags associated with unsafe UGCGs. By analyzing self-reported stories, we first compiled keywords related to unsafe UGCGs.
{Utilizing these keywords, we constructed an initial hashtag list,
including terms such as \textit{\#RobloxUGC}, \textit{\#RobloxCondo}, \textit{\#RobloxSex}, \textit{\#RobloxR34}, \textit{\#RobloxKiller}, and \textit{\#RobloxMurderMystery}. 
During the data collection phase, we systematically incorporated any new hashtag associated with a tweet containing one of the predetermined keywords into our evolving list. This list was continuously updated with emerging hashtags from new tweets until a saturation point was reached where no additional unique hashtags were discovered. The comprehensive list of these hashtags is detailed in Appendix~\ref{app:data1}.}

\noindent \textbf{Image Collection.} We used the Official X Streaming API\footnote{https://developer.twitter.com/en/docs/twitter-api} to collect public tweets during the period from January 1, 2020, to December 31, 2022 (\emph{i.e.}, 2 years) based on the hashtags.
More specifically, we used the tool's flag `image' to make sure all the posts we collected were along with images. 
%
%
At the end of our data collection process,  
{we have retrieved 38,182 image-based tweets and extracted 29,858 images immediately between January 2020 and December 2022.}

\noindent \textbf{Image Filtering and Annotation.} 
We selected a random sample of 4,000 images to serve as the subjects of our study. Each image in the dataset was initially processed to enhance its clarity and quality. 
Inspired by the methodology outlined by Phan et al.~\cite{phan2022lspd}, we employed a three-step filtering process to further refine our dataset.
In the first step, we utilized the Python library called Pillow~\cite{pillow2023} to automatically eliminate images with either a width or height of fewer than 300 pixels. 
In the second step, we manually inspected the remaining images, filtering out those not associated with UGCGs. In the third step, we rigorously reviewed the images to ensure each was sufficiently clear to be easily perceived by humans.
Finally, the images were independently annotated by three authors. 
%
%
We utilized Natural Language Processing (NLP) techniques, particularly the agglomerative hierarchical clustering (AHC) method~\cite{müllner2011modern}, to categorize the stories into unsafe groups, resulting in the identification of four primary categories of unsafe UGCGs: ``sexually explicit'', ``violent'', ``bullying'', and ``scam''. By reviewing each group and their representative stories, we built our codebook for UGCG image annotation (see Appendix \ref{app:codebook}). 
We further employed Fleiss' Kappa score~\cite{fleiss1971measuring}, a statistical measure used to assess reliability among multiple annotators.
The scores revealed a progressive improvement in inter-annotator agreement throughout the three coding rounds, beginning with a \textit{substantial} agreement level of 74\%, then increasing to an \textit{almost perfect} agreement level of 92\%, and finally reaching an \textit{almost perfect} agreement level of 100\%.
%
Ultimately, this process resulted in the identification of 2,924 valid UGCG images, including 1,621 images classified as ``sexually explicit'', 202 as ``violent'', and 1,101 as ``safe''. However, we did not find any images that could be categorized as ``bullying'' or ``scam''.

{\subsection{Detection Challenges of Illicit Promotional Images of UGCGs}}

{\subsubsection{Nature of UGCG Promotional Images}}
{
In the course of our data collection and annotation process, we noted a clear distinction in the nature of promotional images for UGCG advertisements compared to conventional game promotions, which typically use renditions crafted by professional entities. 
Specifically, we noted that the images for UGCG advertisements are predominantly generated by individual users who tend to utilize direct screenshots from the games to showcase their content, as exemplified in Figure~\ref{fig:sample}.} 

{To validate this observation and support our hypothesis regarding the nature of UGCG promotion, we conducted an analysis on a randomly selected subset of 500 images from our dataset. 
The findings were significant: 97.8\% of these images were indeed screenshots directly taken from user generated games. 
This prevalence underscores the subtlety with which these advertisements integrate into the platform, highlighting the need for enhanced moderation tools, which are essential to identify and mitigate the discreet yet pervasive spread of unsafe UGCG advertisements, safeguarding users from potentially unsafe content.}

\begin{figure}[t]
    \centering
    \includegraphics[width=\columnwidth]{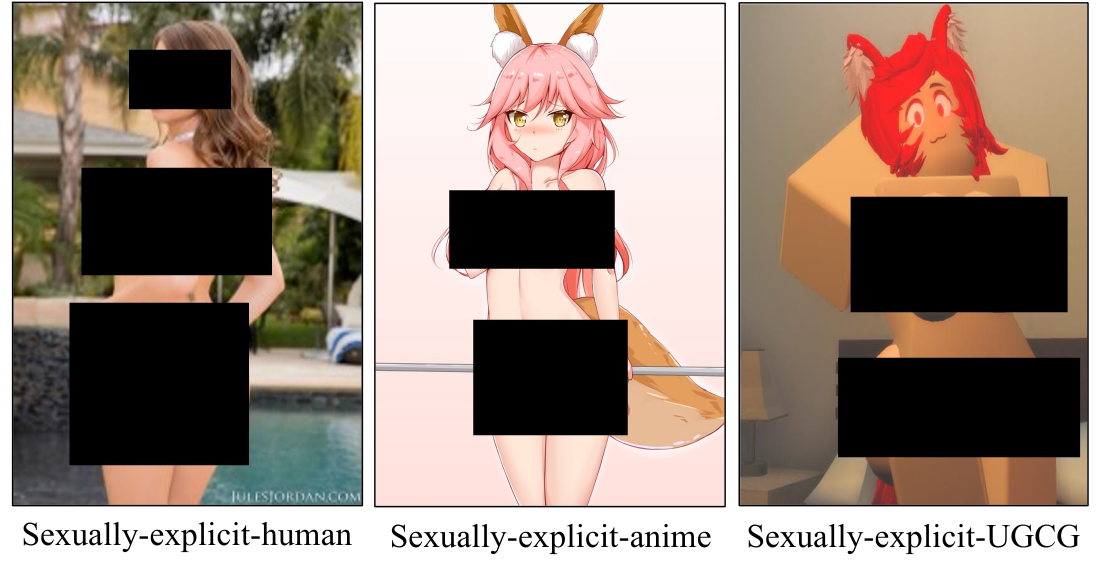}
    \caption{{Samples of sexually explicit images. The Google Vision API's prediction for the image as \textit{sexually-explicit-human}: ``VERY\_LIKELY'' for ``Adult'' and ``Racy''; for the image as \textit{sexually-explicit-anime}: ``VERY\_LIKELY'' for ``Adult'', ``POSSIBLE'' for ``Racy''; for the image as \textit{sexually-explicit-UGCG}: ``UNLIKELY'' for ``Adult'' and ``Racy''.}}
    \label{fig:sex_image}
\end{figure}

\vspace{3mm}
\subsubsection{Evading State-of-the-Art Unsafe Image Detectors}
\label{Sec:SOTA}

Following our previous studies, We wanted to investigate the feasibility of using existing detectors for moderating the image-based illicit online promotion of UGCGs.
To understand the effectiveness of existing unsafe image detectors, we conducted an experiment regarding state-of-the-art (SOTA), commercially available detectors by measuring their capability to detect promotional UGCG images from our dataset.
In our work, we selected five SOTA detectors that are widely used and have the capability to detect unsafe images, which are Clarifai~\cite{clarifai2023}, Yahoo Open Not Safe For Work (NSFW)~\cite{yahoo2016nsfw}, Amazon Rekognition~\cite{amazon2023rekognition}, Microsoft Azure~\cite{microsoft2023azure}, and Google Vision AI~\cite{google2023vision}. 
Due to the ubiquity and effectiveness of these detectors, they can be considered representative of the technology used to defend against unsafe content in existing online platforms. 
To study the capability of these detectors about new and traditional types of sexually explicit images, our experiments incorporated an existing dataset~\cite{kim2021nsfw} that encompasses traditional sexually explicit content, both from real-world scenarios and animated sources. From this dataset, we randomly selected 1,000 images each from the categories labeled ``porn'' and ``hentai'', representing real-world (\emph{i.e.}, sexually-explicit-human) and animated explicit content (\emph{i.e.}, sexually-explicit-anime), respectively. 
Concurrently, another 1,000 images were randomly selected from our annotated dataset, all labeled as ``sexually-explicit'' (\emph{i.e.}, sexually-explicit-UGCG). This selection served to evaluate the detectors' proficiency in identifying sexually explicit content in UGCGs, offering a comparative insight into their capability to detect traditional and emerging forms of explicit imagery. Figure~\ref{fig:sex_image} is an example demonstrating the different categories of images used in this study. 

The responses from existing detectors exhibited a diversity in their output formats. Systems like Clarifai, Yahoo Open NSFW, and Amazon Rekognition provided probability scores in the range from 0 to 1 as outputs to quantify the likelihood of images being unsafe. In contrast, Microsoft Azure presented a binary true or false label to categorize such images.
Google SafeSearch offered a more nuanced classification, with labels ranging from ``UNKNOWN'' to ``VERY\_UNLIKELY'', ``UNLIKELY'', ``POSSIBLE'', ``LIKELY'', and ``VERY\_LIKELY'', each indicating the varying degrees of probability for an image being deemed unsafe. Given this diversity in output formats, establishing a unified criteria for evaluation was crucial.
Based on these varying methods of measuring whether or not an image is unsafe, we used the following thresholds to determine if an unsafe image is detected. 
For Clarifai, Amazon Rekognition, Azure, and Yahoo Open NSFW are logit values in the range 0$-$1. The threshold value is set at 0.5 for unsafe images.
For Google Vision AI, if the model output is ``LIKELY'' or ``VERY\_LIKELY'', we assume that the model identifies unsafe images, which is a common criterion that has been discussed in previous work~\cite{291025}.

\begin{table}[b!]
\centering
\resizebox{\columnwidth}{!}{
\setlength\tabcolsep{0.7ex}
\begin{tabular}{llllll}
\toprule
\multicolumn{1}{c}{\multirow{5}{*}{\textbf{Image Type}}}           & \multicolumn{5}{c}{\textbf{State-of-the-Art Unsafe Image Detectors}}\\
\cmidrule{2-6}
\multicolumn{1}{c}{}                                               & \textbf{\begin{tabular}[c]{@{}l@{}}Clarify \end{tabular}} & \textbf{\begin{tabular}[c]{@{}l@{}}Yahoo\\ Open\\ NSFW\end{tabular}} & \textbf{\begin{tabular}[c]{@{}l@{}}Amazon\\ Rekog-\\ nition\end{tabular}} & \textbf{\begin{tabular}[c]{@{}l@{}}Micro-\\ soft\\ Azure\end{tabular}} & \textbf{\begin{tabular}[c]{@{}l@{}}Google \\ Vision \\AI\end{tabular}} \\
\midrule
\begin{tabular}[c]{@{}l@{}}Sexually-explicit\\ -human\end{tabular} & 88\%                                                                &  92\%                                                                           &  98\%                                                                                &   92\%                                                                            & 98\%                                                                              \\
\begin{tabular}[c]{@{}l@{}}Sexually-explicit\\ -anime\end{tabular} &  89\%                                                               & 81\%                                                                            &  91\%                                                                                &  90\%                                                                             & 99\%                                                                              \\
\rowcolor[HTML]{E8E8E8} 
\begin{tabular}[c]{@{}l@{}}Sexually-explicit\\ -UGCG\end{tabular}  & 13\%                                                                &  13\%                                                                          &  17\%                                                                               &    15\%                                                                           & 67\%\\
\bottomrule
\end{tabular}
        }
        \caption{Effectiveness of SOTA unsafe image detectors.}
        \label{tab:sota}
    \end{table}

\begin{figure*}[t]
    \centering
    \includegraphics[width=\textwidth]{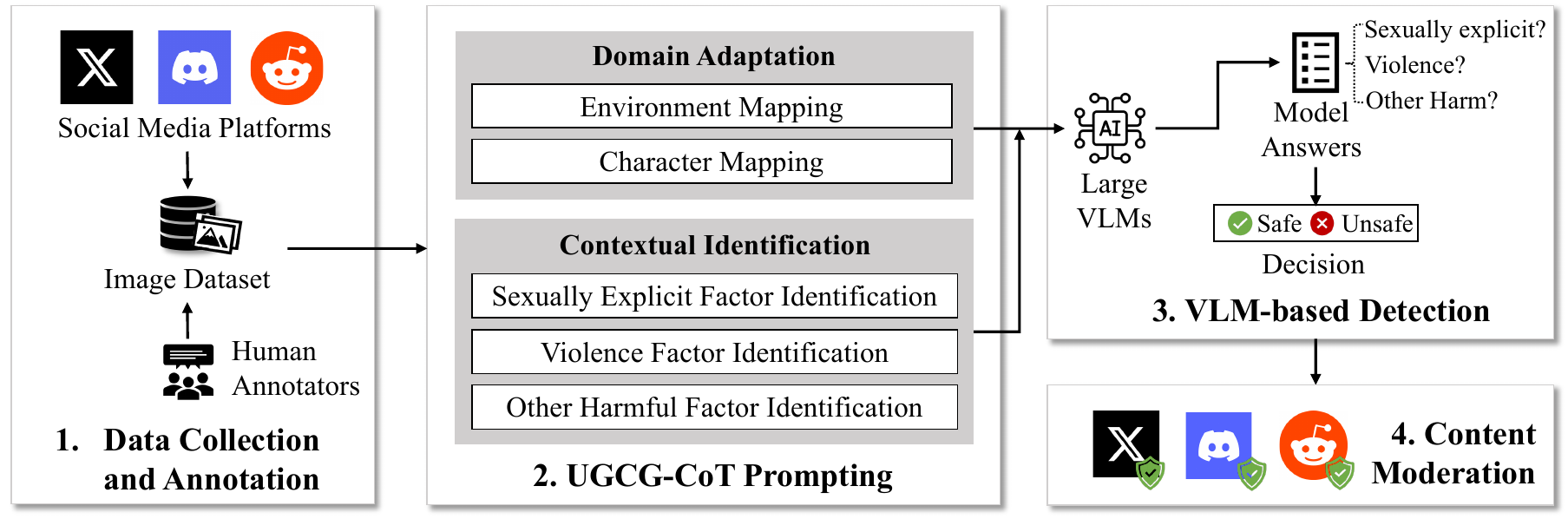}
    \caption{Overview of \ProjectName{}.}
    \label{fig:framework}
\end{figure*}

For the selected unsafe image detector, Google Vision AI, we provide different types of unsafe images as inputs and observe the detection results of the detector. 
Table~\ref{tab:sota} presents the results of this experiment. 
The results indicate that SOTA exhibit exemplary performance in detecting unsafe images depicting humans and anime characters, with all detectors achieving an average detection rate exceeding 90\%. Notably, Google Vision AI demonstrates superior efficacy, achieving detection accuracies of 98\% and 99\% for human and anime sexually explicit images, respectively.
Contrastingly, a pronounced decline in performance is observed when these models are tasked with identifying unsafe content within UGCG images. The detection efficacy of these sophisticated tools is markedly compromised in this context, with four of the tested models yielding detection rates below 20\%. Google Vision AI, albeit the most effective among the evaluated models with a detection rate of 67\%, falls short of the accuracy and reliability requisite for real-world content moderation applications.
We believe that the superior performance of both human and anime images could be attributed to the similarity in the domain of these images, and also to the presence of training samples from both domains. Anime images, although animated, are semantically close to real human images, and are defined similarly to real humans. However, this is not the case for the UGCG images. As a result, this comparative analysis underscores the challenges in the detection of unsafe content within UGCGs: there is a marked domain shift in the case of UGCG images that renders existing systems severely limited.\looseness=-1



\section{\ProjectName{} Design}

\subsection{Overview of \ProjectName{}}
{The overview of our framework, \ProjectName{},
is presented in Figure~\ref{fig:framework}, which consists of four main components: 
(1) Data Collection and Annotation;
(2) \PromptName{} Prompting; 
(3) VLM-based Detection; and 
(4) Content Moderation.}
{The framework begins by compiling a dataset of illicit online promotional images for UGCGs, utilizing the methodology described in Section~\ref{sec:data2}.
\ProjectName{} incorporates 
human annotators to verify and label images based on the activities identified from our study of unsafe UGCGs.}
Following this, we develop \PromptName{} prompts, a novel Chain-of-Thought~(CoT) reasoning-based prompting strategy tailored to enable reasoning-based decision-making for the identification of images used for the illicit promotion of unsafe UGCGs, by addressing the challenges of domain shift and contextual reasoning posed by these images via conditional prompting and {reasoning-based prompting}.
In the next stage, a large VLM is leveraged to run the prepared \PromptName{} prompts on a post with illicit promotional images, and
the output from the VLM is parsed for answers to each of
the \PromptName{} prompts. 
Finally, in our concluding stage, we use the parsed answer from the previous stage to determine whether the post contains illicit promotional images of unsafe UGCGs, and if yes, the post is flagged for moderation.


\subsection{Our Approach}

\subsubsection{Crafting \PromptName{} Prompts}
\label{sec:prompt}
Navigating the intricacies of UGCGs requires our prompts to be adept at tackling the dual challenges of domain adaptation and contextual identification by leveraging large VLMs to make moderation decisions effectively. However, VLMs cannot be directly used for this purpose. 
In the following, we discuss how our novel 
{CoT reasoning-based prompting approach, \PromptName{}, {which is illustrated in Figure~\ref{fig:prompts}.
This approach combines} conditional prompting and reasoning-based prompting strategies and can effectively leverage VLMs to perform domain adaptation and contextual identification, respectively.}


\noindent \textbf{Conditional Prompting for Domain Adaptation.} 
Domain adaption is a crucial methodology that facilitates AI/ML models in navigating and adapting to the transition between distinct domains, enhancing their performance and applicability across varied contexts~\cite{8103149, Csurka2017}.
In this light, our domain-adaptation prompt is instrumental in aiding the model to delineate the nature of the provided images. 
We utilize conditional prompting for domain adaptation to instruct \ProjectName{} to understand the specific characteristics of UGCG images in a zero-shot manner.
{More specifically, the purpose of conditioning in our approach is to refine the search space of VLMs, thereby enhancing the overall effectiveness of our approach. In our setup, we employed prompting as a method of conditioning, concentrating exclusively on images akin to those found in UGCGs. This strategy also aids in minimizing the influence of irrelevant training data on the model.} 
This refinement is achieved through a two-step structure consisting of a condition and a guidance question.
The condition articulated as $Condition$: \textit{``This is an image generated from a role-playing game.''} serves to anchor the model's understanding, clarifying that the image in question is a simulation, not a real-world photograph. 
This foundational insight is pivotal in ensuring the model's responses are contextually anchored.
Complementing this, the guidance question $Q_1$: \textit{``Are there any characters or avatars in this image?"} directs the model's attention to identify human-like figures within the UGCG images. 
This dual structure ensures that the model is not only informed of the simulated nature of the images but is also guided to focus on specific elements within them, facilitating a more refined and contextually appropriate analysis.

\begin{figure}[t]
    \centering
    \includegraphics[width=\columnwidth]{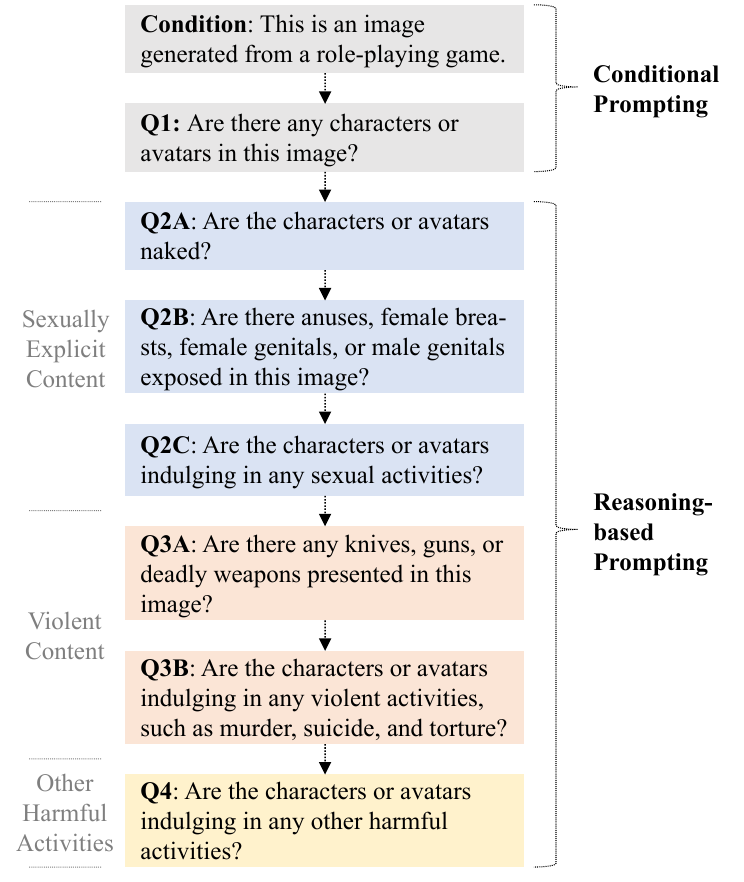}
    \caption{\PromptName{} prompting.}
    \label{fig:prompts}
\end{figure}

\noindent {\textbf{Reasoning-based Prompting for Contexual Identification.} Having equipped large VLMs with the capability to interpret the distinct domain of UGCGs, we proceed to introduce reasoning-based prompts tailored to identify specific unsafe content categories, including sexually explicit material and violence. 
Such prompts enhance \ProjectName{} ability to perform reasoning, consequently enabling the VLM to make contextual decisions~\cite{wei2023chainofthought} and ensuring that unsafe content can be identified and moderated.}
Our image filtering and annotation processes revealed a crucial observation: despite the pronounced domain shift characteristic of UGCG images, the attributes defining unsafe content remain analogous to those in traditional real-world images. Drawing inspiration from established methodologies in unsafe real-world image detection~\cite{tian2018color, phan2022lspd, basch2020assessment, wikipedia2023graphic}, we meticulously crafted our contextual prompts.
We pose the following question prompts to identify sexually explicit content: 
$Q_{2A}$: \textit{``Are the characters or avatars naked?''}; 
$Q_{2B}$: \textit{``Are there anuses, female breasts, female genitals, or male genitals exposed in this image?''}; 
$Q_{2C}$: \textit{``Are the characters or avatars indulging in any sexual activities?''}
To detect violent content, we incorporated these questions:
$Q_{3A}$: \textit{``Are there any knives, guns, or  deadly weapons presented in this image?''}; 
$Q_{3B}$: \textit{``Are the characters or avatars indulging in any violent activities, such as murder, suicide, and torture?''}
We further exploit the analytical prowess of large VLMs with an additional question to uncover a broader spectrum of harmful activities:
$Q_4$: \textit{``Are the characters or avatars indulging in any other harmful activities?''} 
In the end, 
through our comprehensive Contextual Identification reasoning prompts, we equip large VLMs with the enhanced capability to efficiently identify and flag unsafe UGCG images, ensuring a balanced approach that is both domain-specific and context-sensitive.

\subsubsection{Leveraging Large VLM for Processing \PromptName{} Prompts}

We leverage VLMs to process the UGCG images in conjunction with the \PromptName{} prompts, with specific criteria to ensure optimal performance.
The first requirement for running our prompts is that the selected VLM should be comprehensively trained on an extensive array of varied vision-language tasks. Training on both vision and language modalities allows VLMs to understand the commonalities among various input domains, thus giving them domain adaptation capabilities. Furthermore, the incorporation of powerful encoders ensures the precise extraction and processing of complex factors and features inherent in UGCG images.
Our second requirement stipulates that the model must possess enhanced reasoning capabilities. This is paramount due to the intricate nature of UGCG images and the nuanced, potentially unsafe content embedded within them that needs contextual detection. 
The adoption of the chain-of-thought methodology~\cite{wei2023chainofthought} ensures advanced reasoning capacity. This approach has been empirically validated to significantly elevate the decision-making proficiency of large language models (LLMs)~\cite{wei2023chainofthought}. 
In this scenario, the role of LLMs becomes pivotal. 
They act as decision-making models that process vision and language, facilitating reasoning-based decision-making based on both textual and visual features.
Moreover, LLMs, enriched by their extensive training datasets, are essential for embodying CoT reasoning, helping \ProjectName{} leverage the extensive knowledge based on these models to make accurate decisions.
Thus, the synergistic combination of large VLMs, intensively trained across a spectrum of tasks and endowed with amplified reasoning faculties, emerges as a suitable choice to run our prompts.

Our system leverages a VLM to operationalize our prompting strategy in the following way. 
Given image input $X_{vision}$  and supplemented by the \PromptName{} prompts, characterized as the language input $X_{language} \in \{X_{language}^{condition}, X_{language}^{Q_1}, X_{language}^{Q_{2A}}, ...\}$, 
the output is computed as,
\begin{equation} 
\label{eq:vlmoutput}
    \hat{y} = \text{argmax }p(y|X_{vision}, X_{language}).
\end{equation}
Using our \PromptName{} prompts, we decompose the primary problem of detection of unsafe images used in the illicit promotion of unsafe UGCGs into a series of sub-problems. This enables a structured and sequential approach to decision-making, where the final output $\hat{y}$ is a culmination of insights derived from intermediate states. 
To be specific, the steps are as follows.

\noindent \textbf{Step 1.} 
{We first condition the VLM to enable domain adaptation from the real-world context to the simulated gaming environment, enhancing its ability to interpret UGCG images. 
In this process, the model integrates and processes the UGCG image, denoted as $X$, in conjunction with a specified condition $C$. This integration tailors the model's attention features, represented as $X_C$, ensuring they are attuned to the nuances of the UGCG content within the image.}

\noindent \textbf{Step 2.} We then condition the VLM to achieve domain adaptation from identifying real humans to identifying human-like figures within the UGCG image, depicted as follows:
\vspace{-0.5mm}
\begin{equation}
\label{eq:guidence}
    A_1 =  \text{argmax }p(a|X_C, Q_1),
\end{equation}
where $a$ is an intermediate answer that the VLM could output, such as $Yes$, $No$, $N/A$, etc.
In this step, \ProjectName{} identifies valid characters. If detected, the framework proceeds to perform a context check; otherwise, it allows the input image to pass through unchanged without further evaluation.

\noindent \textbf{Step 3.} Next, we prompt the VLM to check if sexually explicit content is presented in the input image.
\vspace{-0.5mm}
\begin{equation}
\label{eq:sex}
    \begin{split}
        A_{2a} &=  \text{argmax }p(b|X_C, Q_{2A}),\\
        A_{2b} &=  \text{argmax }p(c|X_C, Q_{2B}),\\
        A_{2c} &=  \text{argmax }p(d|X_C, Q_{2C}),
    \end{split}
\end{equation}
where $b, c, d,...$ are intermediate answers as $a$ in Equaiton~\ref{eq:guidence}.

\noindent \textbf{Step 4.} Then, we prompt the VLM to identify if violent content is presented in the input image.
\vspace{-0.5mm}
\begin{equation}
\label{eq:violence}
    \begin{split}
        A_{3a} &=  \text{argmax }p(e|X_C, Q_{3A}),\\
        A_{3b} &=  \text{argmax }p(f|X_C, Q_{3B}).\\
    \end{split}
\end{equation}

\noindent \textbf{Step 5.} In addition, we prompt the VLM to output whether other harmful activities are shown in the image.
\begin{equation}
\label{eq:other}
    A_{4} =  \text{argmax }p(g|X_C, Q_{4}).
\end{equation}

\noindent \textbf{Step 6.} The final decision regarding the safety of a UGCG image is determined by aggregating all previous outputs from the VLM. If any of the contextual questions receive a positive answer \emph{i.e.}, yes, the UGCG image is flagged as unsafe. This decision is represented as follows: 
\vspace{-0.5mm}
\begin{equation} \label{eq:incitation}
  \begin{split}
    A_{\text{all}} &= \{A_{2a}, A_{2b}, A_{2c}, A_{3c}, A_4\}, \\
    \hat{y} &= 
    \begin{cases} 
      \text{unsafe}, & \text{if }\exists  a_i \in A_{\text{all}} : a_i = \text{yes}, \\
      \text{safe}, & \text{otherwise}.
    \end{cases}
  \end{split}
\end{equation}

\begin{algorithm}[b!]
    \caption{Illicit UGCG Promotional Image Safety Analysis Using Large VLM}
    \label{alg:cot}
    \SetAlgoLined
    \DontPrintSemicolon
    \SetNoFillComment

    {\bfseries Input:} UGCG image $X_{vision}$, \PromptName{} prompts $T$ $=$ $\{C$, $Q_{\text{dom}}$, $Q_{\text{sex}}$, $Q_{\text{violent}}$, $Q_{\text{harm}}$, $Q_{\text{dec}}\}$, Inference Function $F$, Large Vision Language Model (M);\\
    {\bfseries Output:} Safety prediction $\hat{y}$ with reason\;
    
    \tcp*{Domain Adaptation}
    {$M_{X_C} \gets M(X,C)$\;}
    $y_1 \gets F_{\text{domain}}(M_{X_C}, Q_{\text{dom}})$\;
    
    \If{$y_1$  = ``no''}{
        \Return $y_1$\;
    } 
    
    \tcp*{Contextual Identification}
    \For{$q \in Q_{\text{sex}} \cup Q_{\text{violent}} \cup Q_{\text{harm}}$}{
        $A_q \gets F(M_{X_C}, q)$\;
        \If{$A_q = \text{``yes''}$}{continue}
    }

    \tcp*{Decision Making}
    \If{any $A_q = \text{``yes''}$}{
        $\hat{y}$  = \text{`unsafe'}\\
        \Return $\hat{y}$, \text{Flag the image and issue a warning with the reason.}\;
    }
    \Else{
        $\hat{y}$  = \text{``safe''}\\
        \Return $\hat{y}$, \text{Approve the image for sharing.}\;
    }
\end{algorithm}

\subsubsection{Illicit Online Image Promotion Moderation}
In this section, we outline the deployment of \ProjectName{} for real-world social media platforms, as detailed in Algorithm~\ref{alg:cot}. \ProjectName{} employs the large VLM to evaluate promotional images of UGCGs systematically. Initially, it ascertains the presence of a valid character in the image. Subsequently, the framework assesses the content to identify sexually explicit or violent elements. Additionally, \ProjectName{} scrutinizes the image for other potential safety concerns and annotates any identified issues.
In the end, \ProjectName{} aggregates the responses from the large VLM to determine the safety status of the promotional image. If an image is identified as ``unsafe'', \ProjectName{} will flag the image and issue a warning stating why. Otherwise, \ProjectName{} will approve the image for sharing on social media platforms. The comprehensive assessment ensures that images are flagged appropriately, enhancing the safety and quality of content on social media platforms.


\section{Implementation and Evaluation}
In this section, we first discuss the implementation of multiple components of our system, followed by experiments to evaluate our approaches to identify illicit promotional images of unsafe UGCGs from different perspectives.
Our evaluation goals are summarized below.
\begin{itemize}
    \item Understand the effectiveness of \ProjectName{} by comparing it against existing baseline detectors. ($\S$~~\ref{subsec:compare})
    \item Analyzing the capability of \ProjectName{} to address the challenge of the shift from traditional unsafe image to the UGCG input domain by comparing it with state-of-the-art object detection methods. ($\S$~~\ref{subsec:domain_shift})
    \item Investigating the effectiveness of the conditioning process of \ProjectName{}. ($\S$~~\ref{subsec:condition_promt})
    {\item Examining the effectiveness of the contextual identification process of \ProjectName{}. ($\S$~~\ref{subsec:contextual_promt})}
    \item Running \ProjectName{} on ``in-the-wild''  samples from diverse social media platforms. ($\S$~~\ref{subsec:in-the-wild})
    \item Examining the limitations of traditional vision models for detecting unsafe UGCG images. ($\S$~~\ref{subsec:vision-only})
\end{itemize}


\begin{figure*}[t]
    \centering
    \includegraphics[width=0.95\textwidth]{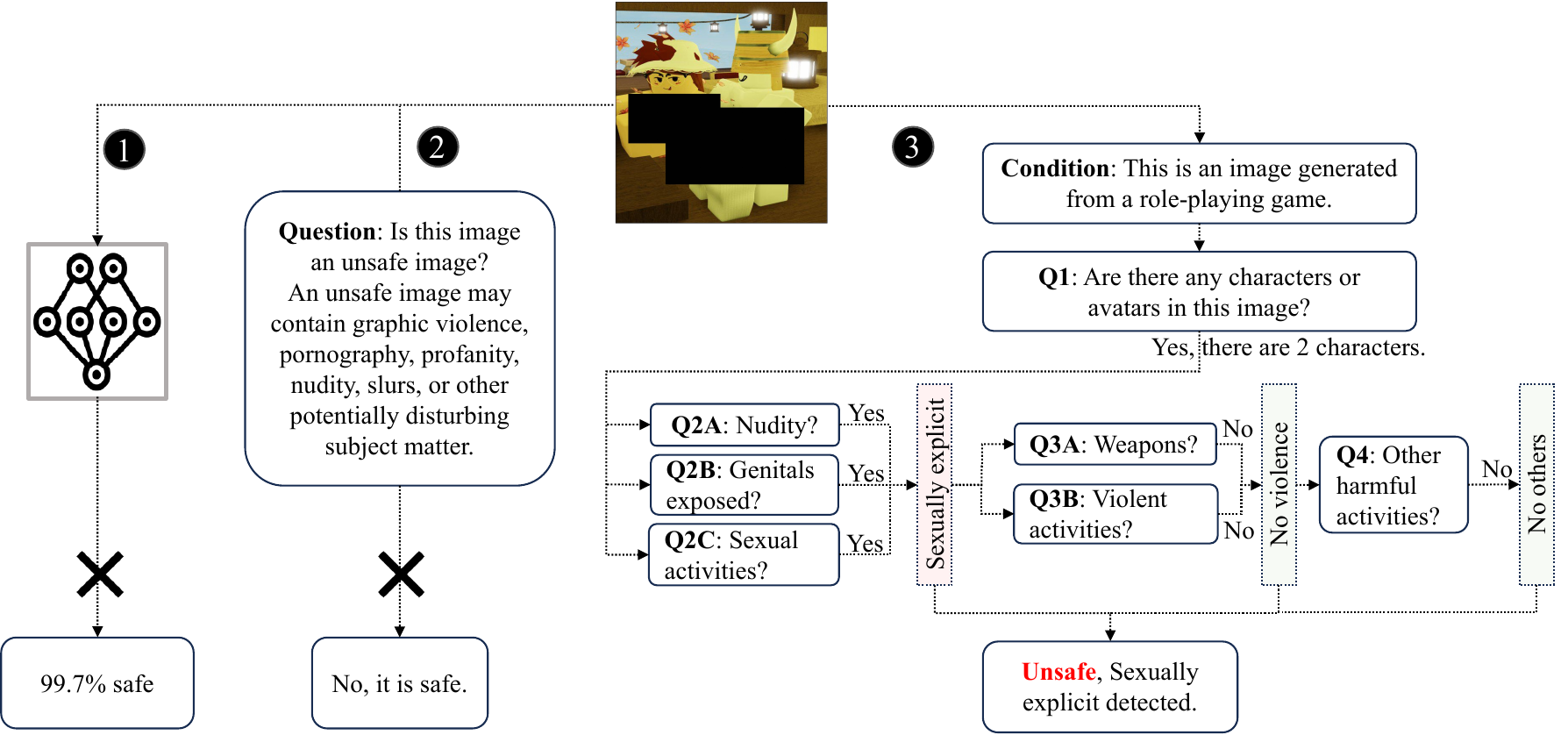}
    \caption{ \protect\inlinegraphics{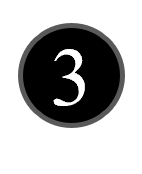}\ProjectName{} for image-based illicit promotion of UGCG detection compared to \protect\inlinegraphics{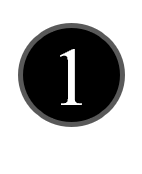}traditional unsafe image detection models and \protect\inlinegraphics{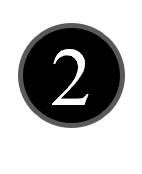}large VLMs with general prompting strategy.}
    \label{fig:approach}
\end{figure*}

\subsection{Implementation Details}
In this section, we discuss the implementation specifics of \ProjectName{}. We utilized the InstructBLIP model, \textit{instructblip-vicuna-13b}~\cite{dai2023instructblip}, as our preferred large VLM for the large-scale execution and evaluation of \PromptName{} prompts. 
We deployed \ProjectName{} with a High-Performance Computing (HPC) system equipped with two 40GB Nvidia A100 graphics cards.
Most of our evaluation experiments were conducted using the labeled dataset described in Section~\ref{sec:data2}, 
{with an additional 322 illicit promotional UGCG images gathered from Reddit and Discord. 
This dataset formed as a test case that represented ``in-the-wild'' UGCG images, providing insights into the model's performance in real-world scenarios, as detailed in Section~\ref{subsec:in-the-wild}. In this experiment, we also built and evaluated \ProjectName{} based on the GPT-4V(ision) model to demonstrate the generalizability of our framework. 
The other specific parameter settings, excluding those set to default values, are discussed in the evaluation.}

\subsection{Baselines}
We evaluated \ProjectName{} with several pertinent baselines to establish a comparative evaluation:
(1) Google Vision AI, a SOTA commercial approach renowned for its efficacy;
(2) A Convolutional Neural Network (CNN) model~\cite{kim2014convolutional}, trained with large-scale image data exceeding 60 gigabytes, comprises a diverse collection of real-world and animated NSFW images renowned for its benchmark performance~\cite{man}. In our experiment, we name it as ``NSFW-CNN''.
(3) A general prompt for unsafe image detection, employing the same large VLM as \ProjectName{}, to evaluate the impact of different prompt strategies on detection performance.

\subsection{Comparison Against Baselines}
\label{subsec:compare}


In this experiment, we evaluated the efficacy of \ProjectName{} in identifying image-based illicit promotion of unsafe UGCGs, comparing its performance against existing baseline detectors for unsafe images. 
We employed Google Vision AI~\cite{google2023vision} as a representative of cutting-edge commercial systems for unsafe image detection.
Additionally, we considered the NSFW-CNN~\cite{man}, a CNN model built upon MobileNetV2 architecture~\cite{sandler2019mobilenetv2}, which is fine-tuned on an extensive dataset of unsafe images to enhance its detection capabilities.
%
Also, to understand the effectiveness of \PromptName{}, we contrasted its performance with the same large VLM model, InstructBLIP, which was probed using a general zero-shot prompting strategy, and our general prompt is crafted by aligning with the widely accepted definition of unsafe content~\cite{wikipedia2023nsfw}, was phrased as follows: 
``Is this image unsafe? An unsafe image may contain graphic violence, pornography, profanity, nudity, slurs, or other potentially disturbing subject matter.''

\begin{table}[b!]
\resizebox{\columnwidth}{!}{
    \begin{tabular}{lcccc}
    \toprule
    \textbf{Defenses}                                                & \textbf{Accuracy} & \textbf{Precision} & \textbf{Recall} & \textbf{F1} \\ 
    \midrule
    
    \begin{tabular}[c]{@{}l@{}}Google \\ Vision AI\end{tabular}     &        0.68           &        0.79            &       0.68          &         0.65    \\ 
    \begin{tabular}[c]{@{}l@{}}NSFW-CNN\end{tabular}        &         0.5         &            0.63        &           0.57      &         0.47    \\ 
    \begin{tabular}[c]{@{}l@{}}InstructBLIP\\ -GEN\end{tabular} &    0.6                &      0.88             &    0.42             &      0.57      \\ 
    \rowcolor[HTML]{E8E8E8}
    \ProjectName{}                                                            &    0.94             &        0.98            &     0.91            &    0.94         \\
    \bottomrule
        \end{tabular}
    }
    \caption{Comparing \ProjectName{} against the baselines.}
    \label{tab:baseline}
\end{table}

The results of our experiment are depicted in Table~\ref{tab:baseline}. 
The NSFW-CNN pre-trained model demonstrates the worst performance, as evidenced by its accuracy of 0.5 and F1 score of 0.47. 
This can be attributed to its high dependency on the dataset it was trained on. The images of UGCGs are very different compared to both real-world and animated visuals, presenting that the traditional pre-trained AI/ML models, such as the NSFW-CNN model, can not be sufficiently equipped to handle due to the training data limitations.
Google Vision AI and InstructBLIP with a general prompt (``InstructBLIP-GEN'') exhibit comparable results, with accuracy levels of 0.68 and 0.6, respectively. 
However, these systems are not suitable for practically flagging such content due to their limited efficacy.
InstructBLIP-GEN excels in the precision score, indicating its strength in minimizing false positives. However, its recall is the lowest, pointing towards the limitation in identifying unsafe images of UGCGs.
Google Vision AI has a more ``balanced'' performance, although it still achieves limited recall and F1 scores.
\ProjectName{} distinctly outperforms all the baselines, with an impressive accuracy of 0.94 and an F1 score of 0.94, indicating its effectiveness in flagging images-based illicit promotion of unsafe UGCGs.
The results underscore the efficacy of employing chain-of-thought reasoning~\cite{wei2023chainofthought} in identifying unsafe UGCGs, as illustrated in the sample depicted in Figure~\ref{fig:approach}. 
Unlike the general zero-shot prompting strategy that probes the large VLM once, \PromptName{} adopts a multi-step decision-making approach, integrating model conditioning and contextual questions. 
This method capitalizes on the large VLM's capabilities to enhance the detection of illicit promotional images of UGCGs through a structured reasoning process. 
It ensures that the final decision is not just an immediate output, but is derived from a comprehensive evaluation, resulting in increased precision and recall.




\subsection{Capabilitiy of \ProjectName{} in Domain Shift}
\label{subsec:domain_shift}

\begin{table}[b!]
\centering
    \resizebox{\columnwidth}{!}{
    \setlength\tabcolsep{0.7ex}
    \begin{tabular}{lccccc}
    \toprule
    \textbf{Image Type} 
    & \textbf{\begin{tabular}[c]{@{}l@{}}Yolo\end{tabular}} 
    & \textbf{\begin{tabular}[c]{@{}l@{}}SSD\end{tabular}} & \textbf{\begin{tabular}[c]{@{}l@{}}Faster\\ RCNN\end{tabular}} 
    & \textbf{\begin{tabular}[c]{@{}l@{}}Google\\Vision AI\end{tabular}} 
    & \textbf{\begin{tabular}[c]{@{}l@{}}\ProjectName{}\end{tabular}} \\
    \midrule
    \begin{tabular}[c]{@{}l@{}}Sexually-exp-\\licit-human\end{tabular}             & 97.4\%        & 98.8\%       & 99.9\%                                                          &   89.6\% &   99.9\%                                                             \\
    \begin{tabular}[c]{@{}l@{}}Sexually-exp-\\licit-anime\end{tabular}          & 97\%          & 99\%   & 99.9\%      & 89.2\%                                                         &      100\%                                                             \\
    \rowcolor[HTML]{E8E8E8} 
    \begin{tabular}[c]{@{}l@{}}Sexually-exp-\\licit-UGCG\end{tabular}       & 3.2\%         & 18.2\%       & 12.4\%                               &      12.8\%                  & 98.2\%         \\
    \bottomrule
    \end{tabular}
    }
    \caption{Comparing SOTA object detection tools.}
    \label{tab:objdetect}
\end{table}
In this experiment, our goal was to evaluate \ProjectName{}'s capability to interpret images of UGCGs, distinguishing them from the realm of traditional unsafe images. 
We focused on the response to question Q2 within \ProjectName{}, detailed in Section~\ref{sec:prompt}. This approach allowed us to examine the adaptability of \ProjectName{} in identifying the presence of characters or avatars amidst the domain shift to UGCG imagery.
We assessed \ProjectName{}'s adaptability to this new domain by comparing it with four renowned object detection tools: Yolo~\cite{yolo}, SSD~\cite{SSD}, Faster RCNN~\cite{ren2016faster}, and Google Vision AI. These tools are capable of detecting objects that they're trained on and assign a confidence score to each prediction. 
In our study, we focused on the object ``person'' due to the nature of our task, in which unsafe images predominantly depict a persona. A threshold of 0.5 was established for the prediction scores to evaluate the performance of each model in identifying valid human-like figures in UGCG images.
Three distinct datasets were employed to scrutinize the efficacy of each method. These include real-world and animated sexually explicit datasets, each containing 1,000 images, as detailed in Section~\ref{Sec:SOTA}, and an additional 1,000 sexually explicit images from our dataset. Each image in this custom dataset has been manually verified to contain human-like figures, ensuring consistency in our evaluation criteria.\looseness=-1

Table~\ref{tab:objdetect} presents the results. 
Each technique demonstrates effectiveness in detecting real-world and anime people, with Faster RCNN and InstructBLIP achieving exemplary performance. However, their proficiency diminishes significantly when applied to UGCG sexually explicit images, with none of the object detection tools surpassing an 18.2\% detection rate.
In contrast, \ProjectName{} is the only effective tool capable of identifying human-like figures in unsafe UGCG images, showing a remarkable 98.2\% detection rate. 
The stark contrast in detection rates points out the significant shift in the input domain when transitioning from real-world and animated images to UGCGs, and the exceptional performance of \ProjectName{} highlights the important role that large VLMs can play in adeptly navigating and overcoming this challenge, 
marking a notable advancement in adapting to and mitigating the complexities introduced by the diverse content in UGCGs.

\subsection{Effectiveness of Conditioning Prompts}
\label{subsec:condition_promt}

\begin{table}[b!]
\centering
\resizebox{0.95\columnwidth}{!}{
\begin{tabular}{ccccc}
\toprule
\textbf{Methods}                                                   & \textbf{Accuracy} & \textbf{Precision} & \textbf{Recall} & \textbf{F1} \\
\midrule
\begin{tabular}[c]{@{}l@{}}\ProjectName{}\\ without\\ conditioning\end{tabular} &    0.79               &         0.91           &     0.74            &    0.82  \\
\rowcolor[HTML]{E8E8E8} 
\begin{tabular}[c]{@{}l@{}}\ProjectName{}\\ with\\ conditioning\end{tabular}    &     0.94              &      0.98              &   0.91              &   0.94         \\
\bottomrule
\end{tabular}
}
\caption{Comparison of \ProjectName{} w/ and w/o the conditioning process.}
\label{tab:condition}
\end{table}

In this experiment, we investigated the impact of condition prompts in \PromptName{}, \emph{i.e.} $Condition$ and {$Q_1$} as discussed in Section~\ref{sec:prompt}.
To be specific, without the conditioning process, \ProjectName{} probes the large VLM directly with contextual identification questions and derives conclusions based solely on their responses. 


The results can be observed in Table~\ref{tab:condition}. 
Without the conditional prompts, \ProjectName{} exhibits notable performance, displaying an accuracy of 0.79 and an F1 score of 0.82, which significantly outperforms the baselines outlined in Table~\ref{tab:baseline}. 
The incorporation of conditioning, however, elevates \ProjectName{}'s performance exponentially. 
The precision escalates to an impressive 0.98, while recall advances to 0.91, denoting a substantial enhancement in identifying and capturing unsafe content of UGCGs.  
{The results highlight the significance of the conditional prompts in narrowing the search space of large VLMs, enabling them to focus on knowledge pertinent to UGCG or similar images. This focus significantly boosts the overall efficacy of \ProjectName{} in detecting unsafe content.}
This enhancement can also potentially be attributed to the combined vision and language training embedded in the large VLMs~\cite{radford2021learning, huggingface2023dive, li2022blip}. 
In contrast, visual models are solely trained on images. 
This dual training paradigm of large VLMs allows for a more comprehensive understanding and interpretation of content, leading to increased accuracy and efficiency in the detection of illicit UGCG images.



                         

\subsection{Effectiveness of Reasoning-based Prompts}
\label{subsec:contextual_promt}


\begin{table}[t]
\centering
\resizebox{0.9\columnwidth}{!}{
\begin{tabular}{llc}
\toprule
\multicolumn{2}{c}{Prompt Ablations}                       & Detection Rate \\
\midrule
& only Q2A          & 86.5\%         \\
                                       & only Q2B          & 56.6\%         \\
                                       & only Q2C          & 42\%           \\
                                       & only Q2A \& Q2B   & 94.6\%         \\
                                       & only Q2A \& Q2C   & 93.3\%         \\
                                       & only Q2B \& Q2C   & 73.3\%         \\
\multirow{-7}{*}{\begin{tabular}[c]{@{}l@{}}Conditional \\ Prompts +\end{tabular}} & \cellcolor[HTML]{E8E8E8}Q2A \& Q2B \& Q2C & \cellcolor[HTML]{E8E8E8}98.2\% \\
\bottomrule 
\end{tabular}}
\caption{{Ablation study for reasoning-based prompts.}}
\label{tab:contextual_detection}
\end{table}

{In this experiment, we assessed the effectiveness of reasoning-based prompts for contextual identification.
To illustrate this, we specifically focused on the application of these prompts in identifying sexually explicit content. 
Note that we ensured the consistency of our approach by maintaining conditional prompts for the domain adaptation process throughout all evaluations.
We then applied specific reasoning-based prompts for identifying sexually explicit content, Q2A: ``Are the characters or avatars
naked?'', Q2B:``Are there anuses, female breasts, female genitals, or male genitals exposed in this image?'', and Q2C: ``Are the characters or avatars indulging in any sexual activities?'' to the same dataset of 1,000 sexually explicit UGCG images as mentioned in Section~\ref{Sec:SOTA}.} 

{
We further conducted an ablation study to evaluate the effectiveness of our reasoning-based prompting strategy.
As shown in Table~\ref{tab:contextual_detection}, we initially assessed the detection rates for each sexually explicit contextual identification prompt: Q2A, Q2B, and Q2C, with results of 86.5\%, 56.6\%, and 42\%, respectively. 
These results indicate that while nudity is common in sexually explicit UGCG images, reliance on a single prompt for detection is inadequate.
Subsequently, we evaluated the impact of removing each prompt individually. Removing Q2A decreased the detection rate from 98.2\% to 73.3\%. Eliminating Q2B made the detection rate drop to 93.3\%, and upon removing Q2C, the rate decreased to 94.6\%.
The results demonstrate the remarkable effectiveness of \ProjectName{}'s reasoning-based prompts. Together, these prompts establish a comprehensive reasoning process, effectively guiding large VLMs toward accurate final decisions.}


\subsection{Running \ProjectName{} on  Unlabeled Samples ``In-the-Wild''}
\label{subsec:in-the-wild}

We conducted an experiment on the \emph{unlabeled} samples
from two other social media platforms, Reddit and Discord, to simulate an ``in-the-wild'' running scenario that leverages our approach to control the real-world image-based illicit promotion of UGCGs.
On Reddit, we employed the keywords associated with unsafe UGCGs, as identified in our preceding study detailed in Section~\ref{sec:data2}, and manually collected {a total of 112 images used for the illicit promotion of UGCGs. Of these, 33 were classified as safe, while 79 contained unsafe content.}
In the case of Discord, we obtained illicit promotional images of UGCGs utilizing a server listing platform~\cite{topgg2023roblox}. 
This platform enabled our entry into a variety of Roblox game servers on Discord, leading to the discovery of {210 instances for image-based illicit promotions of UGCGs. Among these, 92 images were classified as unsafe, while the remaining 118 were safe.}
Subsequently, we evaluated \ProjectName{} with these in-the-wild unsafe UGCG images and underwent an evaluative process, benchmarked against three pre-existing models: Google Vision AI~\cite{google2023vision}, Clarifai~\cite{clarifai2023}, and NSFW-CNN~\cite{man}.
The recent unveiling of GPT-4V(ision)~\cite{openai2023gptv, openai2022chatgpt} has garnered significant attention, demonstrating exceptional performance in various studies~\cite{roboflow2023gpt4vision, indishmarketer2023gpt4vision}. 
Our framework, \ProjectName{}, with its adaptable architecture, can also be deployed based on different large VLMs such as GPT-4V. 
In this experiment, we assessed the generalizability of \ProjectName{} by integrating it with both InstructBLIP and GPT-4V models.}

\begin{table}[b!]
\centering
\resizebox{\columnwidth}{!}{
\setlength\tabcolsep{0.75ex}
\begin{tabular}{lcccccccc}
\toprule
\multirow{2}{*}{} \multirow{2}{*}{\textbf{Detectors}} & \multicolumn{2}{c}{\textbf{Accuracy}} & \multicolumn{2}{c}{\textbf{Precision}} & \multicolumn{2}{c}{\textbf{Recall}} & \multicolumn{2}{c}{\textbf{F1}} \\
\cmidrule(lr){2-3}
\cmidrule(lr){4-5}
\cmidrule(lr){6-7}
\cmidrule(lr){8-9}
 & R       & D       & R        & D        & R      & D      & R    & D    \\
\midrule
Clarifai & 0.44 & 0.73 & 1 & 1 & 0.22 & 0.27 & 0.36 & 0.43\\
NSFW-CNN & 0.57 & 0.78 & 1 & 1 & 0.4 & 0.41 & 0.57 & 0.58\\
\begin{tabular}[c]{@{}l@{}}Google\\ Vision AI\end{tabular} & 0.71 & 0.87 & 0.98 & 0.96 & 0.59 & 0.74 & 0.74 & 0.83 \\
\rowcolor[HTML]{E8E8E8} 
\begin{tabular}[c]{@{}l@{}}\ProjectName{}\\ 
(InstructBLIP)\end{tabular}
& 0.91 & 0.93 & 0.96 & 0.88 & 0.92 & 0.98 & 0.94 & 0.92\\
\rowcolor[HTML]{E8E8E8} 
\begin{tabular}[c]{@{}l@{}}\ProjectName{}\\ (GPT-4V)\end{tabular}       & 0.88 & 0.9 & 1 & 0.97 & 0.83 & 0.79 & 0.91 & 0.88 \\
\bottomrule
\end{tabular}}
\caption{{The ``in-the-wild'' experiment.}}
\label{tab:in-the-wild}
\vspace{-3mm}
\end{table}

Table~\ref{tab:in-the-wild} illustrates the results of the ``in-the-wild'' experiment. 
In this table, ``R'' stands for Reddit, ``D'' stands for Discord.
Clarifai underperforms on both Reddit and Discord datasets, with F1 scores significantly lower than 0.5, indicating its limited capability in identifying unsafe UGCG images effectively. 
Compared to other baselines, NSFW-CNN achieved better performance with an average F1 score of 0.58. However, it still falls short of the feasibility required for practical applications.
Despite Google Vision AI achieving comparatively decent results on both datasets with an average F1 score of 0.78, its low recall, especially for UGCG images from Reddit, suggests that this popular commercial detector is currently insufficient to meet the challenges posed by such illicit online image promotions. 
%
Both InstructBLIP-based \ProjectName{} and GPT-4V-based \ProjectName{} achieved significantly better results, outperforming other baselines. 
\ProjectName{} integrated with InstructBLIP achieved an impressive average accuracy of 0.92 and an average F1 score of 0.93 across the Reddit and Discord datasets, highlighting its exceptional ability to identify unsafe UGCG images.
Similarly, \ProjectName{} integrated with GPT-4V also showed strong results, with an average accuracy of 0.89 and an F1 score of 0.9, confirming its effective adaptability across different large VLMs.
The overall performance, with an average F1 score of 0.91, validates the possibility of \ProjectName{} for real-world deployment, and it also underscores the generalizability in detecting unsafe UGCG images within various resources.\looseness=-1

\subsection{Comparison Against Traditional Vision Models}
\label{subsec:vision-only}

\begin{figure}[t!]
    \centering
    \includegraphics[width=0.65\columnwidth]{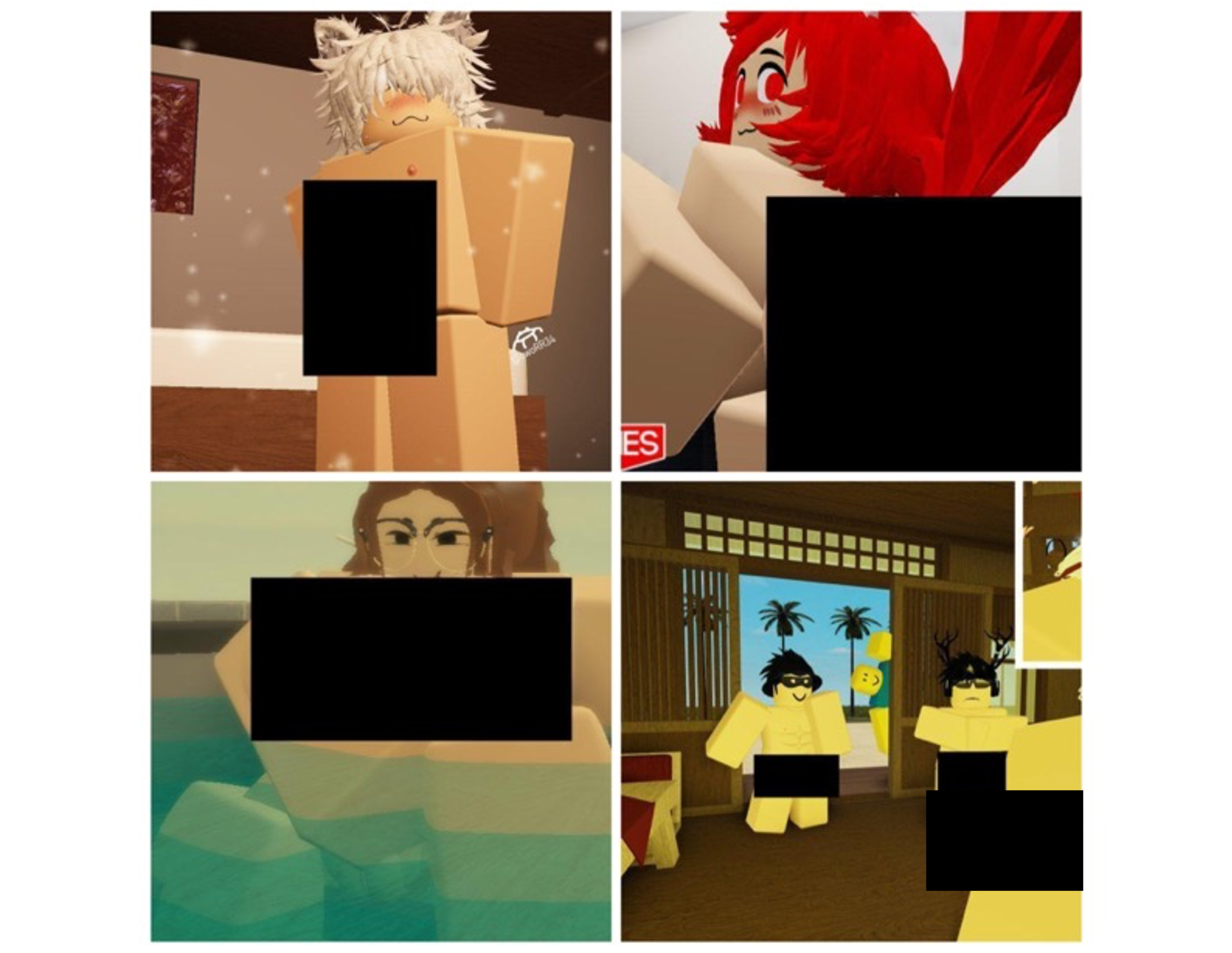}
    \caption{Samples of sexually explicit images used in the ablation study.} 
    \label{fig:ablation}
\end{figure}
In this experiment, we study the limitations of traditional, vision models for the detection of unsafe UGCG images. We fine-tune a ResNet-based CNN model~\cite{he2015deep} utilizing 80\% of the samples from our annotated UGCG image dataset and the rest for testing. 
We conducted an ablation study, where we removed the sensitive regions of 20 unsafe UGCG, as depicted in Figure~\ref{fig:ablation}, and analyzed them with both the ResNet model and our system.
These images should ideally not be considered sensitive since sensitive regions in these images have been removed. This enables us to discern if the ResNet model was effectively identifying unsafe UGCG images contextually, thereby ensuring its effectiveness in detecting such content reliably, or whether they are training on unrelated visual artifacts, like skin color patterns in sensitive images.

Initially, we tested the ResNet model on unaltered samples, achieving 0.9 accuracy and a 0.87 F1 score, competitive with \ProjectName{}. Subsequently, we processed 20 modified images through our system, accurately identifying 18 as ``safe'' due to the removal of explicit content. 
However, the CNN model labeled all modified images as ``unsafe''. 
These outcomes underscore a critical observation: the traditional ResNet model, although adept in certain contexts, tends to overfit, becoming overly sensitive to specific visual cues that are not directly associated with unsafe content and is not practically suitable due to a high false positive rate. 
It highlights an inherent limitation in its capacity to distinguish between genuine unsafe elements and incidental visual patterns. This revelation underscores the necessity for a more refined and nuanced architecture like \ProjectName{} that is adept at detecting unsafe images in a contextual manner. 
\section{Discussion}

\noindent \textbf{Limitations.}
Our study has several limitations. 
First, our analysis is based on reviews obtained from Common Sense Media. 
Although this platform offers a rich array of insights from young users and parents, facilitating the identification of four distinct unsafe topics within UGCGs, it exclusively features English content.
This linguistic limitation potentially narrows the scope of uncovered unsafe content in other languages and cultures. 
Broadening the research to incorporate additional review and community platforms, especially those in diverse languages, could yield a more nuanced and comprehensive understanding of the unsafe content prevalent in UGCGs.
{The evaluation of ``in-the-wild'' scenarios was constrained by a manually collected dataset, which has a limited number of data from Reddit and Discord. To enhance the generalizability of \ProjectName{}, efforts will be enhanced towards amassing a broader dataset from various popular social media platforms. It will not only solidify the robustness of \ProjectName{} but also ensure its efficacy and adaptability for real-world digital environments.}
Another limitation is that the focus of our study is currently confined to UGCGs within Roblox. 
Numerous other platforms exist, such as Minecraft~\cite{Minecraft}, Terasology~\cite{terasology2023}, and LEGO Worlds~\cite{legoworlds2023}, which enable users to create their content and attract a significant population of children and adolescents. 
It is plausible that unsafe activities prevalent in Roblox might also be present in these platforms, and potentially manifest distinct characteristics that necessitate specific moderation strategies. 
Therefore, extending the analysis to include these platforms could provide a more holistic understanding of the safety challenges associated with UGCGs.
Additionally, we observed that illicit promotions of unsafe UGCGs occasionally employ diverse modalities, including GIF images and short videos. 
We believe that once we collect enough data for these varied content formats, the incorporation and evaluation could fortify our analysis, offering a more comprehensive assessment of the multifaceted nature of the illicit promotion of unsafe UGCGs.


\noindent \textbf{Ethical Considerations.}
In our work, we annotated the illicit promotional images of unsafe UGCGs by three of the authors, and no additional workers were recruited in the whole study. 
Our data collection task was approved by IRB.
Every author participant in the process understands the unsafe content before our task.
In our paper, we ensured the removal of mentions to user accounts so that no user information could be traced via public social media.
In addition, we will take all the necessary steps, such as only sharing the data with verified researchers.




\noindent \textbf{In-game Content Moderation}
Our present work is centered on identifying and moderating the image-based illicit promotion of unsafe UGCGs, a crucial endeavor given the associated risks posed to online users, particularly children~\cite{kou2023harmful}. 
However, we believe that the insights gleaned from our current research can shed light on the expansion of future work on in-game unsafe content moderation. 
By moderating both the game promotions and UGCGs themselves, we can minimize the dangers posed by the UGCGs as comprehensively as possible, thereby protecting users, especially young individuals, shielding them from exposure to unsafe content and activities.

\section{Conclusion and Future Work}
\label{sec:future_work}

In this work, we have embarked upon an initial journey to understand and detect the threat of illicit promotion of unsafe UGCGs. 
We have conducted
{an insight study to understand the limitation of existing unsafe image detectors and present the urgent need for effective moderating mechanisms.}
With the studies, we proposed a novel framework \ProjectName{} to practically address the problem of image-based illicit UGCG promotions.
Our evaluation demonstrates that \ProjectName{} can effectively capture the unsafe images used in the illicit promotion of unsafe UGCGs.


In the future, we plan to extend our framework to adapt it for moderating not only image-based illicit promotions but also in-game unsafe content. 
%
Furthermore, the evolving landscape of Virtual Reality (VR) presents both novel opportunities and challenges. 
We envision an extension of our work into the VR domain, adapting and enhancing our methodologies to address the special and complex safety challenges that arise in these immersive environments. 

\section*{Acknowledgements}
This material is based upon work supported in part by the National Science Foundation (NSF) under Grant No. 2228617, 2129164, 2120369, 2245983, 2237238, 2329704, 2112878 and a National Centers of Academic Excellence in Cybersecurity grant No. H98230-22-1-0307.

\bibliographystyle{unsrt}
\bibliography{bibtexes}



\appendix
\section{Hashtags for UGCG Image Collection}
\label{app:data1}
Here we provide the complete list of hashtags we used for UGCG image data collection in Table~\ref{tab:hashtags}.

\begin{table}[h]
\centering
\resizebox{\columnwidth}{!}{
\begin{tabularx}{9.5cm}{X}
\toprule
Hashtags\\
\midrule
\#RobloxUGC, \#RobloxCondo, \#RobloxSex \#RobloxR34, \#RobloxR63, \#RobloxF**k, \#RobloxPorn,
\#rr34, \#rr63, \#rosex, \#legosex, \#robloxlewd, \#robloxlewdsex, \#robloxnsfw
\#RobloxKiller, \#RobloxBlood, \#RobloxMurderMystery,
\#RobloxMurderMystery2, \#RobloxMM, \#RobloxMM2,
\#RobloxPiggy, \#RobloxWar, \#RobloxWarrior, \#RobloxCombatWarrior, \#RobloxWelcomeToPhantom, \#RobloxWelcome2Phantom, \#RobloxW2P, \#RobloxArenal, \#RobloxBattleGround
\\
\bottomrule
\end{tabularx}
}
\caption{List of hashtags.}
\label{tab:hashtags}
\end{table}

\section{UGCG Image Annotation Codebook}
\label{app:codebook}

{We provide the codebook used for annotating the UGCG image data collections, depicted in Table~\ref{tab:code2}.}

\begin{table}[h]
\centering
\resizebox{\columnwidth}{!}{
\begin{tabular}{l}
\toprule
Codebook for Illicit Promotional Images of UGCGs\\
\midrule
Are there any characters or avatars in this image? \\
\begin{tabular}[c]{@{}l@{}}Are the characters or avatars naked or are any private parts exposed \\ in this image?\end{tabular} \\
\begin{tabular}[c]{@{}l@{}}Are the characters or avatars indulging in any  sexual activities?\end{tabular}                  \\
\begin{tabular}[c]{@{}l@{}}Are any knives, guns, or deadly weapons presented in this image?\end{tabular}                  \\
\begin{tabular}[c]{@{}l@{}}Are the characters or avatars indulging in any violent activities?\end{tabular}                  \\
\begin{tabular}[c]{@{}l@{}}Are the characters or avatars indulging in any bullying activities?\end{tabular}                 \\
Is there any scam content in this image? \\
\begin{tabular}[c]{@{}l@{}}Are the characters or avatars indulging in any other harmful activities?\end{tabular}\\
\bottomrule
\end{tabular}
}
\caption{{Codebook for illicit promotional images of UGCGs.}}
\label{tab:code2}
\end{table}

\appendix
\section{Artifact Appendix}
This artifact appendix is meant to be a self-contained document that
describes a roadmap for evaluating our artifact. 
It should include a

\subsection{Abstract}
{\em In this work, we aim to address the issue of illicit image-based promotions of unsafe user generated content games (UGCGs) on social media. In our study, we collect a real-world dataset comprising 2,924 images that display diverse sexually explicit and violent content used to promote UGCGs by their game creators. We additionally create a cutting-edge system, UGCG-Guard, designed to aid social media platforms in effectively identifying images used for illicit UGCG promotions. This system leverages recently introduced large vision-language models (VLMs) and employs a novel conditional prompting strategy for zero-shot domain adaptation, along with chain-of-thought (CoT) reasoning for contextual identification. UGCG-Guardachieves outstanding results, with an accuracy rate of  94\% in detecting these images used for the illicit promotion of such games in real-world scenarios.}

\subsection{Description \& Requirements}

We have prepared a CSV file that stores the image path of each UGCG image and its annotation label. Meanwhile, we prepared Python 3 scripts to reproduce the results of UGCG-Guard. The evaluator should have a Python 3 environment ready to run the prepared scripts.

\subsubsection{Security, privacy, and ethical concerns}

Our artifact contains an unsafe image dataset of inappropriate content, such as sexually explicit and violent images from Roblox games. Please avoid viewing the images directly if you are not comfortable with them.

\subsubsection{How to access}
The artifact can be accessed via the GitHub link: \url{https://github.com/UBSec/UGCG-Guard/tree/1072d5c51a0e7bae2290da08e957e5b1d86cd7b6}.


\subsubsection{Hardware dependencies}
UGCG-Guard is a framework for integrating large VLMs to detect insecure UGCG images. Our framework can utilize both open-source and closed large VLMs.
In this work, we have prepared two scripts to ensure evaluators can successfully run UGCG-Guard with different environments.

The open-source large LVLM-based UGCG-Guard requires a runtime environment with over 50 GB of total GPU memory.

The closed LVLM-based UGCG-Guard requires only API requests and has no hardware dependencies.


\subsubsection{Software dependencies}
None.

\subsubsection{Benchmarks}
{None.}

\subsection{Set-up}
The Python 3 environment is necessary for our artifacts. In addition, please make sure that ``base64'', ``requests'', and ``pandas'' are installed in your Python 3 environment.

\subsubsection{Installation}
For open-source large LVLM-based UGCG-Guard, we have prepared the ``requirements.txt''. Run the following code in your Python environment to install the necessary dependencies.
\begin{lstlisting}[language=Python]
    pip install -r requirements.txt
\end{lstlisting}

For the closed large UGCG-Guard based on LVLM, make sure you have the OpenAI Python library. If not, you can install it using the following code.
\begin{lstlisting}[language=Python]
    pip install --upgrade openai
\end{lstlisting}


\subsubsection{Basic Test}
The evaluators can run scripts directly for artifact evaluation. They will output an error message if any dependencies are missing.

\subsection{Evaluation workflow}
\subsubsection{Closed large LVLM-based UGCG-Guard}
\begin{enumerate}
    \item Open the script ``gpt.py`` and insert your OpenAI API key by changing the code in line 11:
\vspace{-2mm}
\begin{lstlisting}[language=Python]
    api_key = "YOUR_API_KEY"
\end{lstlisting}
\vspace{-4mm}
    \item Run ``gpt.py``, and the results will be stored in a new CSV file named ``ugcg\_gpt.csv''. Testing the whole dataset will cost around 10 dollars. You can randomly select a subset with a number of images to test by adding a line of code under line 14, such as:
\vspace{-2mm}
    
    \begin{lstlisting}[language=Python]
        df = df.sample(1000)
    \end{lstlisting}
\vspace{-4mm}
    
    \item After the ``ugcg\_gpt.csv`` result file is generated, you can run ``view\_result.py`` to calculate and output the accuracy, precision, recall, and F1-score.
\end{enumerate}

\subsubsection{Open-source large LVLM-based UGCG-Guard}
\begin{enumerate}
    \item Run ``requirements.txt`` to set up your Python environment.

    \item The model we used in this artifact is InstructBLIP from HuggingFace. Note that you may modify the code below to distribute the running task properly to different GPUs.
\begin{lstlisting}[language=Python]
    device_map = infer_auto_device_map(model, max_memory={0: "28GiB", 1: "28GiB"},no_split_module_classes=['InstructBlipVisionModel', 'InstructBlipQFormerModel', 'LlamaDecoderLayer'])
\end{lstlisting}

    \item Run the script ``blip.py''. The results will be automatically stored in ``ugcg\_blip.csv''.
    \item Open ``ugcg\_blip.csv'' and look into the column ``blip\_output'', if any of the answers for Q2-Q7 is ``Yes'', the label should be ``unsafe''. 
    \item Compare the predictions with the ground truth in the ``label'' column. ``1'' indicts unsafe images and ``0'' indicts safe images.
    
\end{enumerate}

\subsubsection{Major Claims}

\begin{compactdesc}

    \item[(C1):] UGCG-Guard achieves a satisfactory performance for detecting unsafe UGCG images. This is proven by
    the experiment described in Section 6.3 in our paper, whose
    results are reported in Table 2.


\end{compactdesc}

\subsubsection{Experiments}
Either of the following experiments can be implemented to reproduce our results.

\begin{compactdesc}

    \item[(E1):] Closed large LVLM-based UGCG-Guard [20 human-minutes + 1.5 compute-hour + 280 MB disk]:

    \begin{asparadesc}

        \item[Preparation:] Please set up your Python 3 environment and ensure the required dependencies are installed. Prepare an OpenAI API key to send requests that use GPT models.

        \item[Execution:]
        Run the ``gpt.py'' file in the Python 3 environment and wait patiently for the CSV file to be generated.

        \item[Results:] \textit{The results are automatically calculated and printed to the terminal. After you execute the ``view\_result.py'' file. The expected output should be around 94\% accurate.}
    \end{asparadesc}

    \item[(E2):] Open-source large LVLM-based UGCG-Guard [1.5 human-hour + 2 compute-hour + 3 GB disk]: 

    \item[Preparation:] Please set up your Python 3 environment and make sure you have the required dependencies installed. The experiment requires a sufficient GPU, otherwise the execution will fail due to lack of memory.

    \item[Execution:]
    Run the ``blip.py'' file in the Python 3 environment and wait patiently for the CSV file to be generated.

    \item[Results:]
    Open ``ugcg\_blip.csv'' with tools such as Microsoft Excel and observe the output stored in the ``blip\_output'' column. If any of the answers for Q2-Q7 is ``Yes'', please mark the predicted label as ``1''. Otherwise, label it as ``0''.
    Compare the predicted label with the records in the ``label'' column. Calculate the accuracy with the following equation:
    \begin{equation*}
        accuracy = \frac{\text{The number of correct predictions}}{\text{The number of total predictions}}
    \end{equation*}
    The expected output should be around 94\% accurate.

\end{compactdesc}





\subsection{Version}
Based on the LaTeX template for Artifact Evaluation V20231005. Submission,
reviewing and badging methodology followed for the evaluation of this artifact
can be found at \url{https://secartifacts.github.io/usenixsec2024/}.

\end{document}